\documentclass[aip,twocolumn,letterpaper]{revtex4}

\usepackage{epsfig}
\usepackage{xcolor}

\usepackage{graphics}
\usepackage{epsfig}

\newcommand{\dn}{\mbox{dn}}
\newcommand{\cn}{\mbox{cn}}
\newcommand{\sn}{\mbox{sn}}

\usepackage[margin=1.0in]{geometry}

\begin{document}
\title{The toroidal flux and separatrix effects in tokamaks} 
\author{Allen H Boozer}
\affiliation{Columbia University, New York, NY  10027 \linebreak ahb17@columbia.edu}

\begin{abstract}

An implication of Faraday's Law is thst the rate of change of the poloidal flux relative to the toroidal flux associated with a magnetic surface is given by the loop voltage.   The emphasis in the tokamak literature on the poloidal almost to the exclusion of the toroidal magnetic flux has led to fundamental mistakes in understanding magnetic field evolution, especially as it affects plasma disuptivity.  Designs of tokamak power plants include a divertor that is defined by a separatrix at the plasma edge.  The use of the toroidal flux clarifies the descriptions of such plasmas. This paper explores the effects of a separatrix on plasma parameters that are used for assessing disruptivity: the edge safety factor and the internal inductance.  The exploration of these effects includes an analytic model, which illustrates why the use of the poloidal instead of the toroidal flux introduces a significant definitional uncertainty in what is meant by the edge safety factor and the internal inductance.

\end{abstract}

\date{\today} 
\maketitle



\section{Introduction}

\begin{figure}
\centerline{ \includegraphics[width=2.5 in]{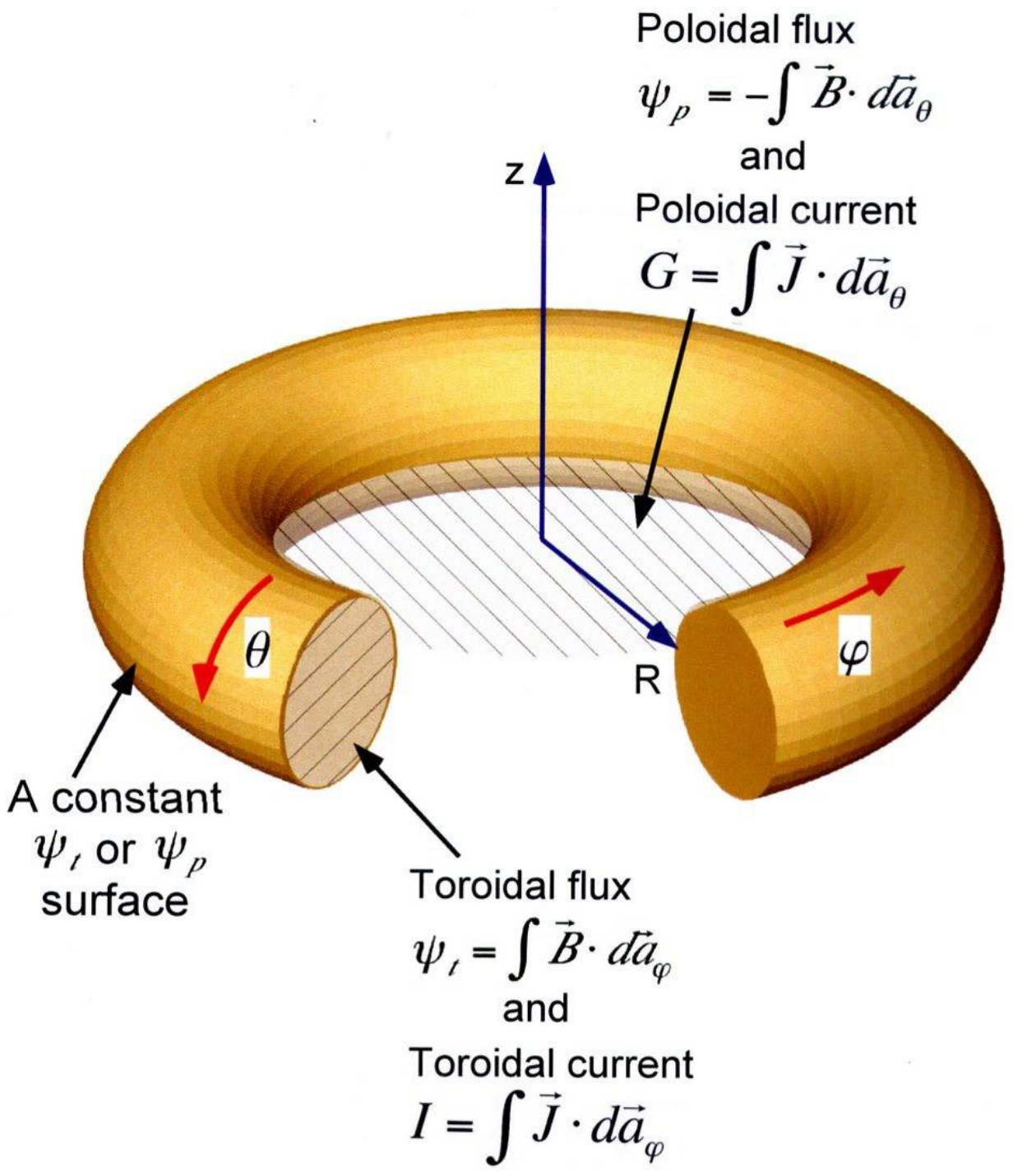}}
\caption{The toroidal flux $\psi_t$ is the magnetic flux enclosed by a toroidal magnetic surface. The poloidal flux $\psi_p(\psi_t,t)$ is  the magnetic flux going down through the hole in the center of the toroidal magnetic surface. The toroidal current is $I(\psi_t,t)$ is the current enclosed by the magnetic surface.  The poloidal current $G(\psi_t,t)$ is the current coming up through the hole in the center of the toroidal magnetic surface.  The current density in the figure is denoted by $\vec{J}$, but in this paper by $\vec{j}$.  This was Figure 1 in Boozer,  Nucl. Fusion \textbf{55}, 025001 (2015).  } 
\label{fig:fluxes-currents}
\end{figure}

The natural radial coordinate in tokamaks and stellarators is a magnetic flux that is associated with each toroidal surface.  For stellarators, the standard choice is the toroidal magnetic flux $\psi_t$ enclosed within each surface.  For tokamaks, the standard choice is $\psi_p^{in}$, normally denoted by $\psi$, which is the poloidal flux that is internal to the plasma and produced by the plasma current.  The flux $\psi_p^{in}$  has a subtle relationship to the poloidal flux $\psi_p$  associated with a toroidal surface, which is the magnetic flux that passes through the central hole of that toroidal surface, Figure \ref{fig:fluxes-currents}.  

The relation between  $\psi_p^{in}$ and the poloidal flux $\psi_p$ is
\begin{equation}
\psi_p^{in} \equiv \Psi_p^{ax}(t) - \psi_p. \label{pol-coord}
\end{equation}
$\Psi_p^{ax}(t)$ is the magnetic flux that passes downward through the circular disk defined by the magnetic axis of a tokamak, and $\psi_p$ is the magnetic flux that passes through the central hole of a toroidal surface, Figure \ref{fig:fluxes-currents}. 
 
Equation (\ref{pol-coord}) for the relation between $\psi_p^{in}$ and $\psi_p$ may seem unnecessarily complicated.   But as explained in the paper \emph{Constraints on the magnetic field evolution in tokamak power plants}  \cite{Boozer:Constraints2026}, the failure to recognize the importance of Equation (\ref{pol-coord}) led to interpreting simple but misleading models of plasma evolution as more rigorous than exact results from Faraday's Law.   This paper is essentially an appendix to  \cite{Boozer:Constraints2026}, which discussed the effect of a separatrix using results derived here. 

Figure \ref{fig:fluxes-currents} illustrates the definition of the toroidal magnetic flux $\psi_t\equiv\int \vec{B}\cdot d\vec{a}_\varphi$ and of the poloidal flux $\psi_p\equiv-\int \vec{B}\cdot d\vec{a}_\theta$.  Reference \cite{Boozer:RMP} proves, using only Faraday's Law and mathematics, that during any interval of time in which a particular magnetic surface persists that encloses the toroidal magnetic flux $\psi_t$, the poloidal flux $\psi_p$ evolves as
\begin{eqnarray}
\frac{\partial \psi_p(\psi_t,t)}{\partial t} &=& V_\ell(\psi_t,t),  \mbox{    where   } \label{flux-ev} \\
V_\ell(\psi_t,t) &\equiv& \lim_{L\rightarrow\infty} \frac{\int_{-L}^L \vec{E}\cdot d\vec{\ell}}{\int_{-L}^L \vec{\nabla}\left(\frac{\varphi}{2\pi}\right)\cdot d\vec{\ell}}
\end{eqnarray}
and $d\vec{\ell}$ is the differential distance along a magnetic field line.   The loop voltage $V_\ell$ gives the slippage of the poloidal relative to the toroidal magnetic flux.  

Equation (\ref{flux-ev}) for the slippage of the poloidal flux relative to the toroidal requires the retention of the toroidal flux $\psi_t$ in an analysis of the evolution of the poloidal field.   Although the proof of Equation (\ref{flux-ev}) is non-triivial, the importance of the definition $\psi_p^{in} \equiv \Psi_p^{ax}(t) - \psi_p$ is trivially proven at  $\psi_p^{in}=0$, which is the magnetic axis of a tokamak.  Faraday's Law plus Stokes' Theorem imply the magnetic flux enclosed by the magnetic axis is 
\begin{equation}
\frac{d\Psi_p^{ax}}{dt} = \oint_{ax} \vec{E}\cdot d\vec{\ell}
\end{equation} 
with $d\vec{\ell}$ the differential distance along the circular magnetic axis.  

\begin{figure}
\centerline{ \includegraphics[width=2.0 in]{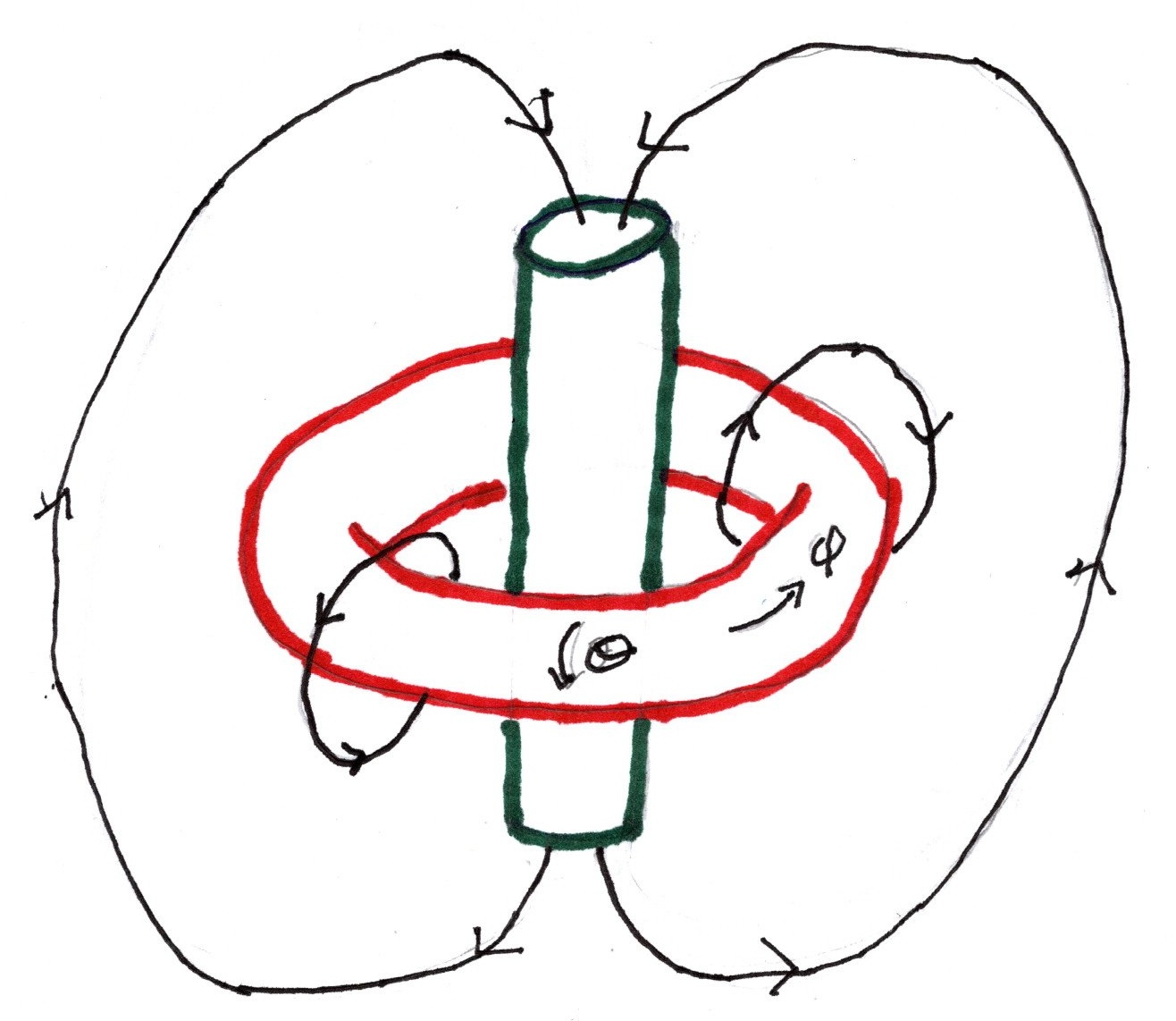}}
\caption{The lines of the poloidal magnetic field produced by the toroidal plasma current are shown together with the magnetic field produced by the central solenoid of a tokamak.  This was Figure 2 in  Phys. Plasmas \textbf{33}, ? (2026); doi:10.1063/5.0322714.}
\label{fig: B}
\end{figure}

As illustrated by Figure \ref{fig: B}, the poloidal magnetic flux enclosed by the magnetic axis is the sum of two terms, 
\begin{equation}
\Psi_p^{ax}(t)=\Psi_p^{sol}(t) + \Psi_p^{pl}(t), \end{equation}
where $\Psi_p^{sol}(t)$ is the flux of magnetic field going downward through the central solenoid and $\Psi_p^{pl}(t)$ is the poloidal flux produced by the current in the plasma.   The poloidal flux produced by the plasma current is itself the sum of two terms, 
\begin{equation}
\Psi_p^{pl}(t) = - \Big(\Psi_p^{in}(t) + \Psi_p^{ex}(t) \Big),
\end{equation}
where $\Psi_p^{in}(t)$ is $\psi_p^{in}$ evaluated at the plasma edge and $\Psi_p^{ex}(t)$ is the flux produced by the plasma current that lies outside the plasma.  Normally \cite{Boozer:Constraints2026}, $\Psi_p^{in}$ and $\Psi_p^{ex}$ are positive with $\Psi_p^{ex}>\Psi_p^{in}$. 

It has long been known \cite{MHD stab} that disruptions occur when the profile of the plasma current profile is too peaked, which gives tearing modes, or insufficiently peaked, which gives external kinks.   A measure of the peakedness of the current profile is the internal inductance $\ell_i$ for poloidal field energy, Equation (\ref{ell_i}).  The feasibility of tokamak power plants depends on adequate disruption avoidance \cite{Eiditis:2021}, and that requires the internal poloidal flux $\Psi_p^{in}$ lie in the range $\Delta\Psi_p^{in}$  that is consistent with acceptable current profiles.  

Unfortunately, $\Delta\Psi_p^{in}/\Psi_p^{pl}<<1$, so careful control, which is easier in tokamak experiments than in power plants, seems required.  In particular, the loop voltage $V_{\ell}$ must be essentially constant across the plasma \cite{Boozer:Constraints2026}.   In the absence of strong current drive or bootstrap current, the current profile is determined by the profile of the plasma resistivity, $\eta$.  The primary dependencies of $\eta$ are the electron temperature and $Z_{eff}$, which is determined by the plasma impurities and their state of ionization.  

The power required to maintain the plasma in an economic power plant should be $\lesssim5\%$ of the power passing through the plasma, but in experiments it is the total power except for the internal resistive heating.  The limitation on the power required for maintenance of a fusion power plant implies both the level external control of the temperature profile by heating and of the level of current profile control by external current drive are small  \cite{Boozer:Constraints2026}. 

The rate of change of the flux in the central solenoid, $d\Psi_p^{sol}/dt$, is directly controllable, but the rate of change of the  poloidal flux enclosed by the magnetic axis, $d\Psi_p^{ax}/dt$ is not---certainly not to the accuracy required to avoid disruptions.

The toroidal flux $\psi_t$ has had little prominence in the literature on tokamaks despite Equation (\ref{flux-ev}) requiring its use to precisely determine the evolution of the poloidal magnetic field in toroidal plasmas.  

The use of the toroidal flux $\psi_t$ enclosed by a magnetic surface also leads to simpler and more precise definitions of the internal inductance, $\ell_i$, and of the edge safety factor when the magnetic surfaces deviate far from circular.  Since tokamak plasmas are thought to require a divertor defined by a magnetic separatrix, the simplicity and precision that comes from the use of $\psi_t$ as a radial coordinate extends to equilibrium conditions and not just the requirement that $\psi_t$ be used to determine the implications of Faraday's Law.


The separatrix of a tokamak with a divertor has several distinct effects:  (1) The shaping of the surfaces has a direct effect on the edge safety factor and the internal inductance, which are captured by $\sigma(\psi_t,t)$, a function determined by the shape of the magnetic surfaces, Section \ref{sec:shape}.  (2) The region near the separatrix is generally characterized by a pedestal \cite{Pedestal:2025} that enhances the confinement time by an approximate factor of two through sustaining a jump in the plasma temperature and density.  This jump produces a strong bootstrap current.  The result is a much larger current density at the plasma edge than without a separatrix, which reduces $\ell_i$ for a fixed edge safety factor.  This makes the plasma more stable to tearing and less stable to external kinks.

In TFTR, which had near circular magnetic surfaces bounded by a limiter, the plasma current profile was adequately described by the edge safety factor $q_a$ and the internal inductance $\ell_i$ to determine when disruptions produced by the current profile would occur \cite{MHD stab}.  In JET plasmas \cite{JET-dis:2020}, which were bounded by a separatrix, the edge safety factor and the internal inductance were not adequate.  Nevertheless, the safety factor at the plasma edge gives important information about the plasma disruptivity.  

There is a subtlety.  The safety factor $q$ is infinite on a separatrix, which makes the direct use the value of $q$ on the separatrix of no value in determining disruptivity.  In 1989, T. S. Taylor et al \cite{Def-q_95} showed that related information about the stability of plasmas bounded by a separatrix can be obtained by replacing the edge safety factor of limited plasmas by $q_{95}$.  The definition of $q_{95}$ uses $\psi_p^{in}$;
\begin{eqnarray}
 q_{95} &\equiv& q\Big((\psi_p^{in})_{95}\Big) \mbox{    where   }  \label{Std-q_95}  \\
(\psi_p^{in})_{95} &\equiv & \frac{\psi_p^{in}}{\Psi_p^{in}} = 0.95.
\end{eqnarray}
$\Psi_p^{in}$ is the total poloidal magnetic flux produced by the plasma current that lies inside the separatrix.  Since tokamak power plants are thought to require a divertor defined by a separatrix, $q_{95}$ quickly became such a standard designation of the edge rotational transform that is hard to trace it to the reference by T. S. Taylor et al.

The use of 95\% of the internal poloidal flux between the magnetic axis and the edge to define the edge safety factor gives a definition that is dependent on the current profile throughout the plasma since \cite{Boozer:Constraints2026}
\begin{eqnarray}
\psi_p^{in}(\psi_t) &=& \Psi_p^{ax} +\int_0^{\psi_t}    \frac{d\psi_t}{q(\psi_t)}. \\
&=& \Psi_p^{ax} +\int_0^{\psi_t}    \frac{I(\psi_t)}{\sigma(\psi_t)G(\psi_t)} d\psi_t \label{psi_p}\label{psi_p}
\end{eqnarray}
using Equation (\ref{q-sigma}) for the relationship between the safety factor $q(\psi_t)$ and the shape function $\sigma(\psi_t)$.  The toroidal $I(\psi_t)$ and poloidal  $G(\psi_t)$ currents are defined in Figure \ref{fig:fluxes-currents}. 

The use of a fraction of the toroidal magnetic flux enclosed by the separatrix, $\Psi_t$, would have made the $q_{95}$-like quantity essentially independent of the current profile.  The sensitivity of $q_{95}$ defined by poloidal flux to the breadth of the current profile is illustrated by Figure \ref{fig:psi_95p}, which gives $\psi_t/\Psi_t$ at the location of 95\% of the internal poloidal flux between the magnetic axis and the separatrix, denoted by $\psi_t^{p95}/\Psi_t$, as a function of the width of the current channel.  Figure \ref{fig:Pi_sigma} allows one to determine the magnitude of the shape function $\sigma$ as a function of $\psi_t/\Psi_t$.   The use of the poloidal flux gives an uncertainty in the enhancement of $\sigma$ by the separatrix due to the uncertainty in $\psi_t^{p95}/\Psi_t.$  The uncertainty  is  that $0.89\gtrsim \psi_t^{p95}/\Psi_t \gtrsim 0.81$.  The value of $\sigma$ is uncertain due to the variation in $(\psi_t^{p95}/\Psi_t)\Big\{1+\Pi_\sigma(\psi_t^{p95}/\Psi_t)\Big\}$, which is between 1.05 and 1.33.  This non-trivial definitional uncertainty due to the central current profile could be eliminated by defining the edge by some definite fraction of $\psi_t/\Psi_t$, such as $\psi_t/\Psi_t=0.85$.

The definition of the internal inductance is also affected by the uncertainty in the definition of the plasma edge.  Reference \cite{Boozer:Constraints2026} shows the internal inductance is given by
\begin{eqnarray}
\ell_i &\equiv& \int_0^1\frac { I(s) q(1)}{I(1) q(s) } ds, \label{ell_i}
\end{eqnarray} 
where $s=\psi_t/\psi_t^{p95}$ with $\psi_t^{p95}$ at the place where the poloidal flux reaches 95\% of difference in the poloidal flux between the separatrix and the axis.  


The paper \emph{Constraints on the magnetic field evolution in tokamak power plants} \cite{Boozer:Constraints2026} was focused on the implications of Faraday's Law, as expressed in Equation (\ref{flux-ev}), on plasma disruptivity and maintenance.  The fundamental errors that are made by those who are not familiar with this equation are pointed out and discussed.  This paper has a different focus.  It is study of the shape function $\sigma(\psi_t,t)$ and how it can be calculated.  A simple analytic example of a separatrix is derived that illustrates both the properties $\sigma(\psi_t,t)$ in the presence of a separatrix and how they can be calculated.  An approximate form of these results was used in \cite{Boozer:Constraints2026} to study the effects of a separatrix on the internal inductance and on the safety factor near the plasma edge.  


Section \ref{sec:profile} defines the plasma current profile, which is thought to be the primary cause of disruptions.

Section \ref{sec:shape} is on the shape function, $\sigma(\psi_t,t)$.   As shown in this section, the shape function in tokamaks fully determines the effect of the shape of the magnetic surfaces on the safety factor and the internal inductance.  The shape function was introduced in Section 5.3.4 of Reference \cite{Boozer:NF-review}.  Any deviation of the plasma surface shape from circular increases $\sigma$ and the safety factor.  An example of the effect of elliptical and triangular shaping on $\sigma$ is derived in this section.

Section \ref{sec:model} introduces the two-wire model of a tokamak with a separatrix, Figure \ref{fig:separatrixt}.  This model allows an analytic determination of the shape function $\sigma(\psi_t,t)$ starting with Cartesian coordinates that are periodic in $z$. The magnetic surfaces are represented using elliptic functions.  The required properties of the elliptic functions and integrals are given in Appendix \ref{sec:elliptic}.  Subsection \ref{sec:shape function} derives the shape function in terms of the complete elliptic integral of the first kind, $K(k)$, Equation (\ref{shape-function}).  This subsection also derives the expression for the safety factor $q=1/\iota$, where $\iota$ is the rotational transform, also in terms of $K(k)$.  Subsection \ref{sec:toroidal flux} expresses the toroidal flux $\psi_t(k)$ in terms of $E(k) - (1-k^2)K(k)$ where $E(k)$ is the complete elliptic integral of the second kind, which also determines $\Psi_t$, the toroidal magnetic flux enclosed by the separatrix.   Subsection \ref{gen-sigma} obtains the expression of $\sigma(\psi_t/\Psi_t)$, which leads to Figure \ref{fig:Pi_sigma}.  Subsection \ref{p95 in psi_t} derives the relation between the location of 95\% of $\psi_p^{in}$ between the magnetic axis and the separatrix and $\psi_t/\Psi_t$ as a function of the width of the current channel.  This is illustrated in Figure \ref{fig:psi_95p}.

Section \ref{sec:summary} gives a summary, which is short because the fundamental results are also given in the Introduction.


\section{The Plasma Current Profile \label{sec:profile}}

The profile of the net plasma current is $I(\psi_t,t)/I_p(t)$, where the total plasma current $I_p(t)\equiv I(\Psi_t,t)$, Figure \ref{fig:fluxes-currents}, with $\Psi_t$ the toroidal flux enclosed by the bounding magnetic surface.   It is the current profile that is generally thought to be the primary determinant of whether a plasma is unstable to either tearing or external kink modes.  These are thought to be the major causes of disruptions from internal plasma effects.  

Not all disruptions are caused by lack of control of the current profile---a piece of wall material falling into the plasma will lead to a fast radiative collapse of plasma confinement.  Nevertheless, it is important to identify parameters that can be measured and controlled to ensure the current profile is stable.  In a highly shaped tokamak plasmas, the use of the toroidal rather than the poloidal flux as the radial coordinate simplifies this identification.

The plasma current density $\vec{j}$ along the magnetic field, or more precisely $j_{||}/B\equiv \vec{j}\cdot\vec{B}/B^2$, has two parts.  One part, the Pfirsch-Schl\"uter current, obeys \cite{Boozer:P-S} the equation 
\begin{eqnarray}
\frac{\partial}{\partial \ell} \frac{j_{||}}{B} &=& 4\pi\frac{dp}{d\psi_t} \frac{\partial (1/B)}{\partial \theta_0}, \mbox{   where  }\\ \label{P-S}
\vec{B} &=& \vec{\nabla}\psi_t\times \vec{\nabla} \frac{\theta_0}{2\pi}.
\end{eqnarray}
The Pfirsch-Schl\"uter current is required to make $\vec{j}$ divergence free in the presence of a gradient in the pressure.  The other part is the homogeneous solution to Equation (\ref{P-S}), which is  $\partial I(\psi,t)/\partial \psi_t$ and is called the net plasma current.  $I(\psi_t,t)\equiv\int \vec{j}\cdot d\vec{a}_\varphi$, Figure \ref{fig:fluxes-currents}, is the toroidal current in Amperes enclosed by a magnetic surface.  

The net plasma current is determined by the loop voltage, through the $\eta \vec{j}$ term in an Ohms Law, plus the contributions of the bootstrap current and of current drive.  The danger of neoclassical tearing modes \cite{Hegna-Callen:1994,La Haye:2006}  limit the usage of the bootstrap current in power plant designs, and the large power \cite{Fisch:1987}---comparable to the alpha-heating power for full current drive \cite{Boozer:Constraints2026}---results in current drive having a limited roll in power plants.


\section{The Shape Function \label{sec:shape} }

In axisymmetry, the shape function $\sigma(\psi_t,t)$ fully determines the effect of the shape of the magnetic surfaces on the safety factor and the internal inductance.

The simplest coordinates in which to give the form of the shape function $\sigma(\psi_t,t)$ are the $(\psi_t,\theta,\varphi)$ coordinates, Figure \ref{fig:fluxes-currents}, in which
\begin{eqnarray}
2\pi \vec{B} &=& \vec{\nabla} \psi_t \times \vec{\nabla}\theta + \vec{\nabla} \varphi \times \vec{\nabla}\psi_p(\psi_t,t);  \label{contra-rep}\\
&=& \mu_0 G(\psi_t,t) \vec{\nabla}\varphi +  \mu_0 I(\psi_t,t) \vec{\nabla}\theta + \beta_*\vec{\nabla}\psi_t. \hspace{0.2in} \label{cov-rep}
\end{eqnarray}  
These two two simple forms for $\vec{B}$ exist  \cite{Mag-coord} whenever or wherever there are magnetic surfaces and a plasma equilibrium, $\vec{\nabla}p=\vec{j}\times\vec{B}$.  The first or contravariant representation, Equation (\ref{contra-rep}), makes the divergence-free nature of $\vec{B}(\vec{x})$ manifest.  The second or covariant form, Equation (\ref{cov-rep}), allows a simple determination of $\vec{\nabla}\times\vec{B}$.   Positions are defined using these coordinates as $\vec{x}(\psi_t,\theta,\varphi,t) = X\hat{X} + Y\hat{Y}+Z\hat{Z}$, with the Cartesian coordinates specified as functions of $(\psi_t,\theta,\varphi,t)$.  The mathematics of general coordinates is derived in a two-page appendix to \cite{Boozer:RMP}.  

The coordinates $(\psi_t,\theta,\varphi)$  that have both a simple contravariant, Equation (\ref{contra-rep}), and covariant, Equation (\ref{cov-rep}), representation  \cite{Mag-coord}  are unfamiliar to many working on tokamaks.  They are well known as Boozer coordinates to those working on stellarators since they are fundamental to showing the strong breaking of axisymmetry need not give unacceptable neoclassical confinement as was thought.  Indeed, the degree to which the neoclassical transport can be reduced appears remarkably unlimited \cite{Landreman-Paul,Goodman}.

Reference \cite{Boozer:Constraints2026} proves that in axisymmetric tori, the shape function is
\begin{eqnarray}
\sigma(\psi_t,t)  &\equiv& \frac{(\partial\vec{x}/\partial\theta)^2}{(\partial\vec{x}/\partial\varphi)^2}. 
\end{eqnarray}
An analytic study of the behavior of the shape function $\sigma(\psi_t)$ in the presence of a separatrix is the focus of this paper.  


\subsection{Generalized Coordinates \label{sec:gen-coord} }

Although the shape function has it simplest form in Boozer coordinates, other poloidal and toroidal angles can be defined and will be used in the analytic separatrix example given in Section \ref{sec:model}.

The contravariant form of the magnetic field, Equation (\ref{contra-rep}), is preserved when the toroidal angle preserves its $2\pi$ periodicity but otherwise undergoes a general alteration, $\varphi= \varphi_* + \nu(\psi_t,\theta_*,\varphi_*)$,  and the poloidal angle $\theta$ is changed to $\theta = \theta_* +\iota(\psi_t) \nu $ with $\iota\equiv d\psi_p/d\psi_t$.  Then
\begin{eqnarray}
2\pi \vec{B} &=&  \vec{\nabla}\psi_t \times  \vec{\nabla} \theta_* +  \vec{\nabla} \varphi_* \times \vec{\nabla}\psi_p(\psi_t); \label{gen-contra-B} \\
&=&  \mu_0 G(\psi_t)  \vec{\nabla} \varphi_* + \mu_0 I(\psi_t) \vec{\nabla}\theta_* + \vec{\nabla} \phi  \nonumber\\ && \hspace{0.2in} +\beta_n\vec{\nabla}\psi_t \mbox{   with  }  \label{gen-cov-B}\\
\phi &\equiv& \mu_0(G+\iota I)\nu.
\end{eqnarray}

Using the dual relations of general coordinates \cite{Boozer:NF-review}, Equation (\ref{gen-contra-B}), can be written using the $(\psi_t,\theta_*,\varphi_*)$ Jacobian, $\mathcal{J}_*$, as
\begin{eqnarray}
2\pi \vec{B} &=& \frac{1}{\mathcal{J}_*} \Big( \frac{\partial \vec{x}}{\partial\varphi_*}  +\iota \frac{\partial \vec{x}}{\partial\theta_*} \Big) \label{direct-contra-B} 
\end{eqnarray}
since  $d\psi_p/d\psi_t \equiv \iota$.  The Jacobian can be found by dotting Equation (\ref{gen-contra-B}) with $\vec{\nabla}\varphi_*$ to obtain  $2\pi \vec{B}\cdot\vec{\nabla}\varphi_* =  (\vec{\nabla}\psi_t\times  \vec{\nabla} \theta_*)\cdot\vec{\nabla}\varphi_*$.  Therefore, 
\begin{eqnarray}                                                                                                                                                                                                                                                                                                                                                                                                                                                                                                                                                                                                                                                                                                                                                                                                                                                                                                                                                                                                                                                                                                                                                                                                                                                                                                                                                                                                                                                                                                                                                                                                                                                                                                                                                                                                                                                                                                                                                                                                                                                                                                                                                                                                                                                                                                                                                                                                                                                                                                                                                                                                                                                                                                                                                                                                                                                                                                                                                                                                                                                                                                                                                                                                                                                                                                                                                                                                                                                                                                                                                                                                                                                                                                                                                                                                                                                                                                                                                                                                                                                                                                                                                                                                                                                                                                                                                                                                                                                                                                                                                                                                                                                                                                                                                                                                                                                                                                                                                                                                                                                                                                                                                                                                                                                                                                                                                                                                                                                                                                                                                                                                                                                                                                                                                                                                                                                                                                                                                                                                                                                                                                                                                                                                                                                                                                                                                                                                                                                                                                                                                                                                                                                                                                                                                                                                                                                                                                                                                                                                                                                                                                                                                                                                                                                                                                                                                                                                                                                                                                                                                                                                                                                                                                                                                                                                                                                                                                                                                                                                                                                                                                                                                                                                                                                                                                                                                                                                                                                                                                                                                                                                                                                                                                                                                                                                                                                                                                                                                                                                                                                                                                                                                                                                                                                                                                                                                                                                                                                                                                                                                                                                                                                                                                                                                                                                                                                                                                                                                                                                                                                                                                                                                                                                                                                                                                                                                                                                                                                                                                                                                                                                                                                                                                                                                                                                                                                                                                                                                                                                                                                                                                                                                                                                                                                                                                                                                                                                                                                                                                                                                                                                                                                                                                                                                                                                                                                                                                                                                                                                                                                                                                                                                                                                                                                                                                                                                                                                                                                                                                                                                                                                                                                                                                                                                                                                                                                                                                                                                                                                                                                                                                                                                                                                                                                                                                                                                                                                                                                                                                                                                                                                                                                                                                                                                                                                                                                                                                                                                                                                                                                                                                                                                                                                                                                                                                                                                                                                                                                                                                                                                                                                                                                                                                                                                                                                                                                                                                                                                                                                                                                                                                                                                                                                                                                                                                                                                                                                                                                                                                                                                                                                                                                                                                                                                                                                                                                                                                                                                                                                                                                                                                                                                                                                                                                                                                                                                                                                                                                                                                                                                                                                                                                                                                                                                                                                                                                                                                                                                                                                                                                                                                                                                                                                                                                                                                                                                                                                                                                                                                                                                                                                                                                                                                                                                                                                                                                                                                                                                                                                                                                                                                                                                                                                                                                                                                                                                                                                                                                                                                                                                                                                                                                                                                                                                                                                                                                                                                                                                                                                                                                                                                                                                                                                                                           
\mathcal{J}_* &\equiv& \frac{1}{(\vec{\nabla}\psi_t \times  \vec{\nabla} \theta_*)\cdot\vec{\nabla}\varphi_*} \\
& = & \frac{1}{2\pi \vec{B}\cdot\vec{\nabla}\varphi_*}.
\end{eqnarray}


\subsection{Relation between the Shape Function and the Safety Factor}

The relation between the shape function and the safety factor can be simply derived in $(\psi_t,\theta_*,\varphi_*)$ coordinates.

In axisymmetry, $(\partial\vec{x}/\partial\theta_*)\cdot(\partial\vec{x}/\partial\varphi_*)=0$.  Dotting Equation (\ref{direct-contra-B}) and Equation (\ref{gen-cov-B}) with $\partial\vec{x}/\partial\theta_*$  and with $\partial\vec{x}/\partial\varphi_*$ respectively gives
\begin{eqnarray}
&& \iota \frac{1}{\mathcal{J}_*}  \left(\frac{\partial \vec{x}}{\partial \theta_*} \right)^2 = \mu_0 I(\psi_t) + \frac{\partial \phi}{\partial \theta_*}; \\
&&  \frac{1}{\mathcal{J}_*}   \left(\frac{\partial \vec{x}}{\partial \varphi_*} \right)^2 = \mu_0 G(\psi_t) + \frac{\partial \phi}{\partial \varphi_*}.
\end{eqnarray}
Integrating both equations over $\theta_*$ and $\varphi_*$ through their full $2\pi$ ranges  implies
\begin{eqnarray}
&& \iota\oint  \left(\frac{\partial \vec{x}}{\partial \theta_*} \right)^2 \frac{d\theta_* d\varphi_*}{\mathcal{J}_*}   = \mu_0 I(\psi_t) \oint\frac{d\theta_* d\varphi_*}{\mathcal{J}_*} ; \\
&&\oint \left(\frac{\partial \vec{x}}{\partial \varphi_*} \right)^2 \frac{d\theta_* d\varphi_*}{\mathcal{J}_*}  = \mu_0 G(\psi_t)\oint \frac{d\theta_* d\varphi_*}{\mathcal{J}_*}. \hspace{0.2in}
\end{eqnarray} 

Defining the shape function as
\begin{eqnarray}
&&\sigma \equiv \frac{\oint  \left(\frac{\partial \vec{x}}{\partial \theta_*} \right)^2 \frac{d\theta_* d\varphi_*}{\mathcal{J}_*}}{\oint \left(\frac{\partial \vec{x}}{\partial \varphi_*} \right)^2 \frac{d\theta_* d\varphi_*}{\mathcal{J}_*} }  \hspace{0.2in} \mbox{    implies   } \label{shape-exp} \\
&&\iota \sigma(\psi_t) = \frac{I}{G}.\label{iota-sigma}
\end{eqnarray}

Since the safety factor $q\equiv1/\iota$, \color{black} the shape function on a magnetic surface is simply related to the safety factor \cite{Boozer:Constraints2026}:
\begin{eqnarray}
q(\psi_t,t) &=& \sigma(\psi_t,t) \frac{G(\psi_t,t)}{I(\psi_t,t)},  \label{q-sigma} 
\end{eqnarray}
where $I(\psi_t,t)$ is the toroidal current enclosed by the magnetic surface and $G(\psi_t,t)$ is the poloidal current outside the magnetic surface, the current that comes up through the central hole in a toroidal magnetic surface, Figure \ref{fig:fluxes-currents}.  


\subsection{Shape Function with Elliptical and Triangular Surface Shaping \label{elip-trian}}

Before studying the effect on the shape function of a separatrix, the effect of elliptical and triangular surface shaping will be considered.  For simplicity, the aspect ratio of the torus will be assumed sufficiently large that geometry can be approximated using Cartesian $(x,y,z)$ coordinates with $z=R\varphi_*$ with the toroidal field $B_\varphi=B_z$ a constant.  The system is assumed periodic in the $z$ direction with a wavelength $2\pi R$. Then, 
\begin{eqnarray}
\vec{\nabla}\varphi_*&=& \frac{\hat{z}}{R}\\
\mathcal{J}_* &=& \frac{R}{2\pi B_\varphi} \\
r&\equiv& \sqrt{\frac{\psi_t}{\pi B_\varphi} }.
\end{eqnarray}

\begin{figure}
\centerline{ \includegraphics[width=2 in]{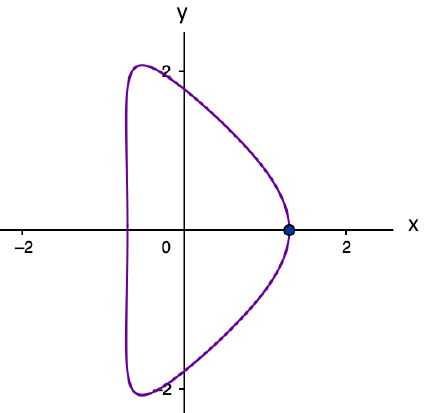}}
\caption{The surface $x = r \cos\theta_* + 0.3(r^2/a) \cos(2\theta_*)$ and $y= 2r \sin\theta_* - 0.3(r^2/a) \sin(2\theta_*)$ at $r=a$ is illustrated.   The $\theta_*=0$ point on the surface is denoted by a dot.}  
\label{fig:plot}
\end{figure}

As a specific example, assume the surfaces are two-to-one vertical ellipses with triangular shaping as well.  The magnitude of the triangular shaping increases by a factor $r/a$ faster than the elliptical shaping.
\begin{eqnarray}
\vec{x}(r,\theta,\varphi) &=& \Big( r \cos\theta_*  + 0.3 \frac{r^2}{a} \cos(2\theta_*) \Big)\hat{x} \nonumber\\ && + \Big(2 r \sin\theta_*  -0.3  \frac{r^2}{a}  \sin(2\theta_*) \Big)\hat{y} \nonumber\\ && + R\varphi \hat{z}.
\end{eqnarray}
Figure \ref{fig:plot} illustrates this surface when $r=a$.  The analyticity of $\vec{x}$ as a function of $x$ and $y$ requires a linear power of $r$ multiply $\cos\theta_*$ and $\sin\theta_*$ and a quadratic power of $r$ multiply $\cos(2\theta_*)$ and $\sin(2\theta_*)$.

\begin{eqnarray}
\Big(\frac{\partial \vec{x}}{\partial\theta}\Big)^2 &=& \Big( r \sin\theta_*  + 0.6 \frac{r^2}{a}\sin(2\theta_*) \Big)^2 \nonumber\\ && + \Big(2 r \cos\theta_*  - 0.6 \frac{r^2}{a} \cos(2\theta_*) \Big)^2 \hspace{0.2in} \\
\Big(\frac{\partial \vec{x}}{\partial\varphi}\Big)^2 &=&R^2. \hspace{0.2in} \mbox{    Consequently,   } \\
\sigma(r) &=& 2.5\frac{r^2}{R^2} + 0.36\frac{r^4}{a^2R^2}. 
\end{eqnarray}

\section{The two-wire model \label{sec:model}}

\begin{figure}
\centerline{ \includegraphics[width=3 in]{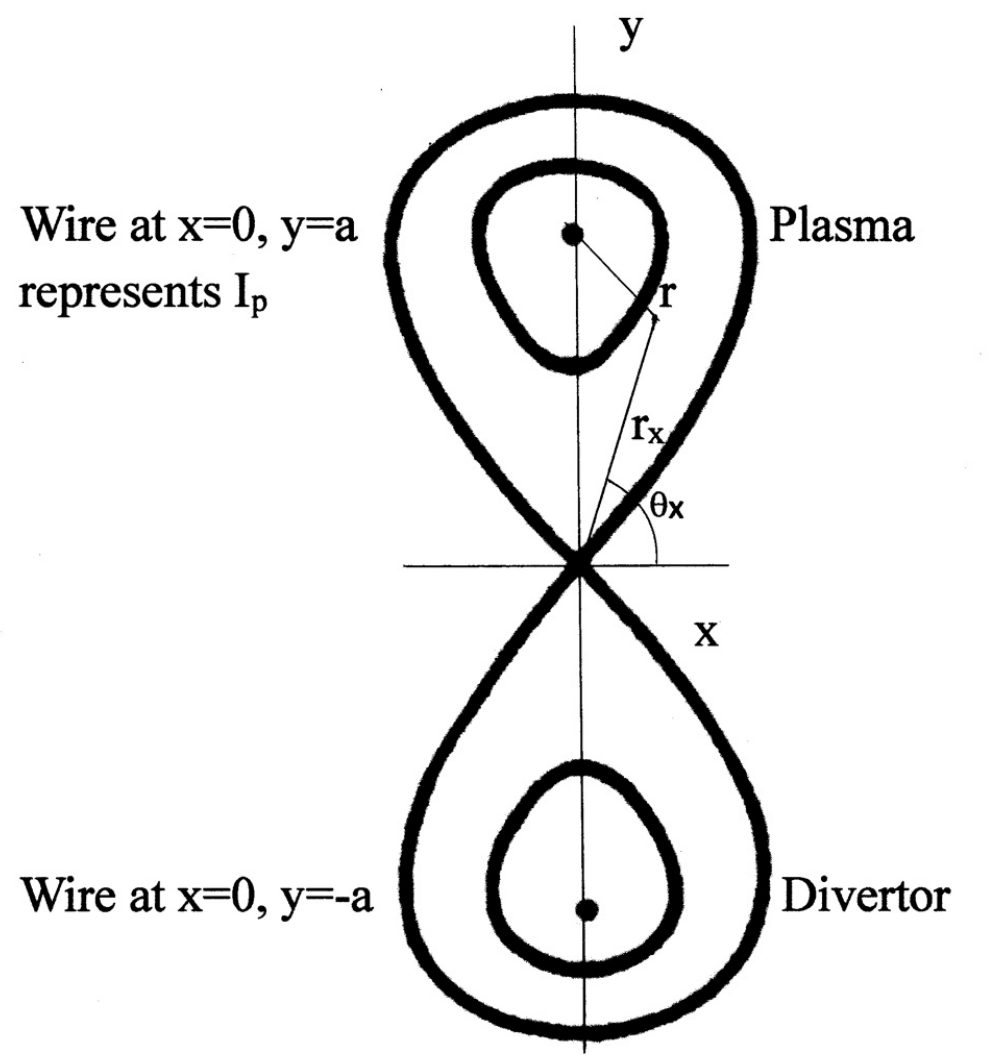}}
\caption{ The two-wire model of a tokamak divertor has a wire at $x=0$, $y=a$ to represent the plasma current $I_p$ and another wire at $x=0$, $y=-a$ carrying a current of equal magnitude and sign to represent the external divertor coil.  Cylindrical coordinates $(r_x,\theta_x,z)$ that have their axis at the X-point of the separatrix are used to define positions, while $r$ gives the distance from the magnetic axis of the plasma---all of the plasma current is assumed to flow in a wire along the magnetic axis.  In order to simplify the figure, the distance from the external divertor coil, $r_e$ is not explicitly illustrated. } 
\label{fig:separatrixt}
\end{figure}

The qualitative features of the shape function in the presence of a separatrix will be illustrated by a simple analytic model for an axisymmetric tokamak with a single-null divertor, Figure \ref{fig:separatrixt}.   The model uses a Cartesian coordinate system in which the $z$ coordinate is assumed to be periodic, $z=R\varphi_*$.  $R$ represents the major radius of the torus, and $\varphi_*$ is the toroidal angle in a $(\psi_t,\theta_*,\varphi_*)$ coordinate system as in Subsections \ref{sec:gen-coord} and \ref{elip-trian}.   

The poloidal magnetic field is assumed to be produced by two parallel wires at $x=0$ and $y =\pm a$ carrying equal currents, which give a divertor-like separatrix with an X-point  at $x=0$, $y=0$.  The top wire is taken to represent the plasma current $I_p$, and the bottom wire the external current that is required to have a divertor, Figure \ref{fig:separatrixt}.    The toroidal field is represented by $B_\varphi \equiv B_z$ and assumed to be a spatial constant.

As the distance $r$ from the wire representing the total plasma current $I_p$ goes to zero, the rotational transform  
\begin{equation}
\iota(r) \rightarrow \frac{\mu_0 R I_p}{2\pi B_\varphi r^2}. \label{close iota}
\end{equation}
The most distant point on the separatrix is at a distance $a$ from the wire that represents the plasma.  Along the separatrix, $\iota=0$, and the safety factor $q\equiv 1/\iota$ is infinite.

In 1968, Hobbs and Taylor \cite{Hobbs-Taylor:1968} discussed the properties of magnetic fields produced by straight wires.  But, the analysis of Boozer and Rechester \cite{B-Rechester:1978} from ten years later provides the mathematical basis for obtaining an analytic solution for the shape function $\sigma(\psi_t,t)$ using using the elliptic functions.  The properties of the elliptic functions are not well known in the fusion community.  The properties that will be used are given in Appendix \ref{sec:elliptic}.

As illustrated in Figure \ref{fig:separatrixt}, the distance from the plasma wire is given by $r^2 = x^2 + (y-a)^2$, and the distance from external wire is given $r_e^2= x^2 + (y+a)^2$.  The distance from the X-point is given by $r_x^2=x^2+y^2$.  Letting $x=r_x\cos\theta_x$ and $y=r_x\sin\theta_x$,
\begin{eqnarray}
r^2 = r_x^2 - 2ar_x \sin\theta_x +a^2; \\
r_e^2 = r_x^2 + 2ar_x \sin\theta_x +a^2.
\end{eqnarray}


\subsection{Poloidal Flux \label{sec:poloidal-flux}}

The magnetic field produced by the wire carrying the plasma current $I_p$ is 
\begin{eqnarray} 
\frac{\mu_0 I_p}{2\pi r}\hat{\theta}  &=& \frac{1}{2\pi} \vec{\nabla} \varphi_*  \times \vec{\nabla}\psi_p^w \\
&=& \frac{\hat{z}}{R} \times\frac{d\psi_p^w}{dr}  \hat{r},  \mbox{   so   }\\
\frac{d\psi^w_p}{dr}  &=& \frac{\mu_0 R I_p}{ r} \\
\psi^w_p(r) & =& \mu_0R I_p \int_a^r\frac{dr}{r} \\
&=& \mu_0R I_p \ln\left(\frac{r}{a}\right).
\end{eqnarray}


 Since both wires contribute the same flux, \color{black} the total poloidal flux is
\begin{eqnarray}
\psi_p^{in} &=& \frac{ \mu_0R I_p}{2} \ln\left(\frac{r^2r^2_e}{a^4} \right) \\
&=& \frac{ \mu_0R I_p}{2} \ln(k^2),  \mbox{  where  }  \label{flux eq}\\
k^2 &\equiv& \frac{r^2r^2_e}{a^4} \\
&=& \Big(\left(\frac{r_x}{a}\right)^2 - 2\left(\frac{r_x}{a}\right)  \sin\theta_x +1\Big)^2\Big(\left(\frac{r_e}{a}\right)^2 \hspace{0.2in} \nonumber\\&&  \hspace{0.2in}+ 2\left(\frac{r_e}{a}\right)  \sin\theta_x +1\Big)^2 \\
&=& \left(\frac{r_x}{a}\right)^4 + \left(\frac{r_x}{a}\right)^2 \Big(2-4\sin^2(\theta_x)\Big) +1\\
&=&  \left(\frac{r_x}{a}\right)^4 + 2\left(\frac{r_x}{a}\right)^2 \cos(2\theta_x) +1,  \label{surf eq}
\end{eqnarray}
since $2\sin^2\theta_x = 1 - \cos(2\theta_x)$.

Since $\vec{B}\cdot \vec{\nabla}\psi_p^{in}=0$, the magnetic surfaces are the constant-$k$ surfaces.  Equation (\ref{surf eq}) gives the equation for the magnetic surfaces, and Equation (\ref{flux eq}) gives the poloidal flux associated with those surfaces.  

Representing the plasma current as lying in a wire makes the poloidal flux $\psi_p^{in}=\mu_0R I_p\ln(k)$ singular.  As $k\rightarrow0$, the poloidal flux goes to minus infinity and $k=2r/a$, which follows from $r_e^2=(2a)^2$ as $r\rightarrow0$.  As the magnetic axis at $r=0$ is approached $\psi_p^{in}=-\mu_0R I_p\ln(a/2r)$; not zero.

A non-singular result for the poloidal flux at the magnetic axis can be obtained if the plasma current density is assumed to be constant out to a radius $r_I<<a$, but zero for larger $r$.  The poloidal flux for $r<<a$  with $\psi_p^{in}(r=0)=0$, as it should be, is
\begin{eqnarray}
\psi_p^{in}(r) &=& 2\pi \mu_0 R \int_0^r B_\theta(r) dr \\
&=&   \mu_0 R I_p \Big(\int_0^{r_I} \frac{r}{r_I^2} dr + \int_{r_I}^r \frac{dr}{r} \Big) \\
&=&  \mu_0 R I_p \Big(\frac{1}{2} + \ln\left(\frac{r}{r_I}\right) \Big) \mbox{   when $r>r_I$   } \hspace{0.2in} \\
&=& \mu_0 R I_p \Big(\frac{1}{2} + \ln\left(\frac{ka}{2r_I}\right) \Big) \\
&=& \mu_0 R I_p \ln\left(\frac{\sqrt{e}k}{k_I}\right), \label{Finite-poloidal}
\end{eqnarray}
where $\psi_p^{in}(k=k_I) = \frac{1}{2}\mu_0 R I_p$ is the smallest value of $\psi_p^{in}$ for which Equation (\ref{Finite-poloidal}) is valid since $k=k_I$ is at the radius at which the current density drops to zero.

Although Equation (\ref{Finite-poloidal}) was derived assuming $k<<1$, a comparison with Equation (\ref{flux eq}) for the poloidal flux shows that it applies for all $k>k_I$.  

In Equation (\ref{flux eq}) an additive constant to the poloidal flux was chosen so the poloidal flux is zero on the separatrix, $k=1$.  Equation (\ref{Finite-poloidal}) implies the poloidal flux between the magnetic axis and the separatrix, 
\begin{eqnarray}
\psi_p^s - \Psi_p^{ax} &=& \mu_0 R I_p \ln\Big(\frac{\sqrt{e}}{k_I}\Big), \mbox{   or } \\
\psi_p^{in}(k) &=& \mu_0 R I_p \ln\Big(\frac{\sqrt{e}k}{k_I}\Big).  \label{k dep of psi_p}
\end{eqnarray}
 The definition of $q_{95}$, Equation (\ref{Std-q_95}), and Equation (\ref{k dep of psi_p}) imply that value of $k$ at 95\% of the internal poloidal flux produced by the plasma current is
 \begin{eqnarray}
 k_{95p} &=& \Big(\frac{k_I}{\sqrt{e}}\Big)^{0.05}  \mbox{   or  } \label{k_I k_95} \\
 k_I &=& \sqrt{e} k_{95p}^{20}.  
 \end{eqnarray}


\subsection{Magnetic surface description using Jacobi elliptic functions \label{sec:surf} }

Equation (\ref{surf eq}) for the shape of the constant-$k$ surfaces, which are the magnetic surfaces, can be represented in terms of the Jacobi elliptic functions by letting
\begin{eqnarray}
\cos(2\theta_x) &=&-\dn(u) \label{theta_x}  \mbox{   and   } \\
\frac{r_x^2}{a^2} &=& \dn(u) + k \cn(u). \label{r_x squared}
\end{eqnarray}

Substituting Equations (\ref{theta_x}) and (\ref{r_x squared}) into Equation (\ref{surf eq}) demonstrates the identity:
\begin{eqnarray}
k^2 &=&  (\dn(u) + k \cn(u))^2 \nonumber\\&& \hspace{0.2in} - 2 \Big(\dn(u) +  k \cn(u)\Big)\dn(u) +1 \\
&=& -\dn^2(u)  +k^2 \cn^2(u) +1\\
&=& k^2 -\Big(k^2 \sn^2(u) + \dn^2(u)\Big) +1, \mbox{  where  } \\
1&=& k^2 \sn^2(u) + \dn^2(u)
\end{eqnarray}
is an identity among the elliptic functions, Equation (\ref{identity 1}) as is $1= \cn^2(u) + \sn^2(u),$ Equation (\ref{identity 0}).

Each magnetic surface in the plasma region is defined by a $k$ with $1\geq k \geq0$.  Poloidal locations on a magnetic surface are given by $u$, with $u$ and $u+4K(k)$ equivalent positions.  That is $u$ has a $4K(k)$ periodicity, which can be turned into the $2\pi$ periodicity of the poloidal angle $\theta_*$ by the identification
\begin{eqnarray}
\theta_* = \frac{2\pi}{4K(k)}u  \label{theta_*}.
\end{eqnarray}
Since $\psi_t$ is a function of $k$ alone, multiplying $\theta_*$ by a $k$ dependent factor does not change the fundamental form of the contravariant, Equation (\ref{gen-contra-B}) representations of $\vec{B}$.  The freedom of $\nu$ implies the fundamental form of the covariant, Equation (\ref{gen-cov-B}), representations of $\vec{B}$ is also unchanged.


\subsection{Shape function calculation \label{sec:shape function}}

The position vector is 
\begin{eqnarray}
\vec{x} &=& r_x \hat{r}_x(\theta_x) + R\varphi_* \hat{z} \\
\frac{\partial \vec{x}}{\partial \theta_*} &=& \left(\frac{\partial r_x}{\partial \theta_*}\right) \hat{r}_x + r_x \left(\frac{\partial \theta_x}{\partial \theta_*}\right) \hat{\theta}_x \\
\left(\frac{\partial \vec{x}}{\partial \theta_*}\right)^2 &=& \left(\frac{\partial r_x}{\partial \theta_*}\right)^2 + r_x^2 \left(\frac{\partial\theta_x}{\partial \theta_*}\right)^2; \\
\left(\frac{\partial \vec{x}}{\partial \varphi_*}\right)^2 &=& R^2.
\end{eqnarray}
The system is assumed to be symmetric in $\varphi_*$, so the partial derivatives with respect to $\theta_*$ can be taken to be total derivatives with respect to $\theta_*$ or $u$.  
\begin{eqnarray}
\sigma &=&\frac{1}{2\pi R^2}\int_0^{2\pi} \left(\frac{\partial \vec{x}}{\partial \theta_*}\right)^2 d\theta_* \\ 
\frac{du}{d\theta_*} &=& \frac{4K}{2\pi},  \mbox{   so   } \\
\sigma &=&\frac{4K}{(2\pi)^2 R^2}\int_0^{4K} \left(\frac{\partial \vec{x}}{\partial u}\right)^2 du\\
&=&4K \frac{\int_0^{4K} \Big(\left(\frac{\partial r_x}{\partial u}\right)^2   + r_x^2 \left(\frac{\partial\theta_x}{\partial u}\right)^2 \Big)du}{(2\pi R)^2}.
\end{eqnarray}

\begin{eqnarray}
 \left(\frac{d r_x}{d u}\right)^2 &=& \frac{1}{4r_x^2}  \left(\frac{d r_x^2}{d u}\right)^2\\
  &=& \frac{a^4}{4r_x^2}  \left(\frac{d(\dn+ k \cn)}{du} \right)^2\\
    &=& \frac{a^4}{4r_x^2}  \left(k\sn(u) \frac{r_x^2}{a^2}\right)^2\\
     &=& \frac{r_x^2}{4}  k^2\sn^2(u).
\end{eqnarray}

Equation (\ref{theta_x}), $\cos(2\theta_x) = -\dn(u)$, can be used to calculate $d\theta_x/du$.  First, note that $d\cos(2\theta_x)/du = -2\sin(2\theta_x) d\theta_x/du$.  Since $\sin^2(2\theta_x) = 1-\cos^2(2\theta_x) = 1- \dn^2(u)$, $\sin(2\theta_x)=\pm k\sn(u)$. The derivative $d\hspace{0.03in} \dn(u)/du=-k^2 \sn(u) \cn(u)$, so $d\theta_x/du =\pm k\cn(u)/2$, and
\begin{eqnarray}
\left(\frac{\partial\theta_x}{\partial u}\right)^2 &=& \frac{k^2 \cn^2(u)}{4}\\
r_x^2\left(\frac{\partial\theta_x}{\partial u}\right)^2&=& \frac{r_x^2}{4}k^2 \cn^2(u),
\end{eqnarray}

\begin{eqnarray}
\left(\frac{\partial \vec{x}}{\partial u}\right)^2 &=&\frac{k^2 r_x^2}{4}\left(\sn^2(u) + \cn^2(u)\right)\\
\sigma &=&\frac{4K}{(2\pi R)^2}\int_0^{4K} \frac{k^2 r_x^2}{4} du\\
&=&\frac{k^2K}{(2\pi R)^2}\int_0^{4K}  r_x^2 du.
\end{eqnarray}
Equation (\ref{r_x squared}) gives $r_x^2$,  $\int_0^{4K} \dn(u) du = 2\pi$,  and $\int_0^{4K}\cn(u)du=0$.  Consequently,
\begin{eqnarray}
\sigma(k)&=& \frac{  k^2K(k)}{2\pi}\frac{a^2}{R^2}.  \label{shape-function}
\end{eqnarray}

When the magnetic surfaces are circular,  $\sigma_c(r) = r^2/R^2$.  Using $r^2 =a^2k^2/4$, which holds for $k<<1$, the circular shape function is 
\begin{eqnarray}
\sigma_c(k)&=& \frac{k^2}{4} \frac{a^2}{R^2} \\
\frac{\sigma(k)}{\sigma_c(k)} &=& \frac{2}{\pi} K(k),
\end{eqnarray} 
which equals unity for $k<<1$.

Using $B_\varphi = \mu_0 G/2\pi R$, the rotational transform
\begin{eqnarray}
\iota &=& \frac{I_p}{G} \frac{1}{\sigma} \\
&=& \frac{I_p}{G} \frac{2\pi R^2}{a^2 k^2 K}\\
&=& \frac{\mu_0 R I_p}{B_\varphi} \frac{1}{a^2 k^2 K}\\
\end{eqnarray}

The rotational transform in the limit $k\rightarrow0$ is $\iota = (\mu_0 R I_p/2\pi r^2)/B_\varphi = RB_\theta/rB_\varphi$, which correctly reproduces Equation (\ref{close iota}).


\subsection{Toroidal flux $\psi_t(k)$ and $\sigma(\psi_t/\Psi_t)$ Calculations \label{sec:toroidal flux}}

\begin{figure}
\centerline{ \includegraphics[width=3 in]{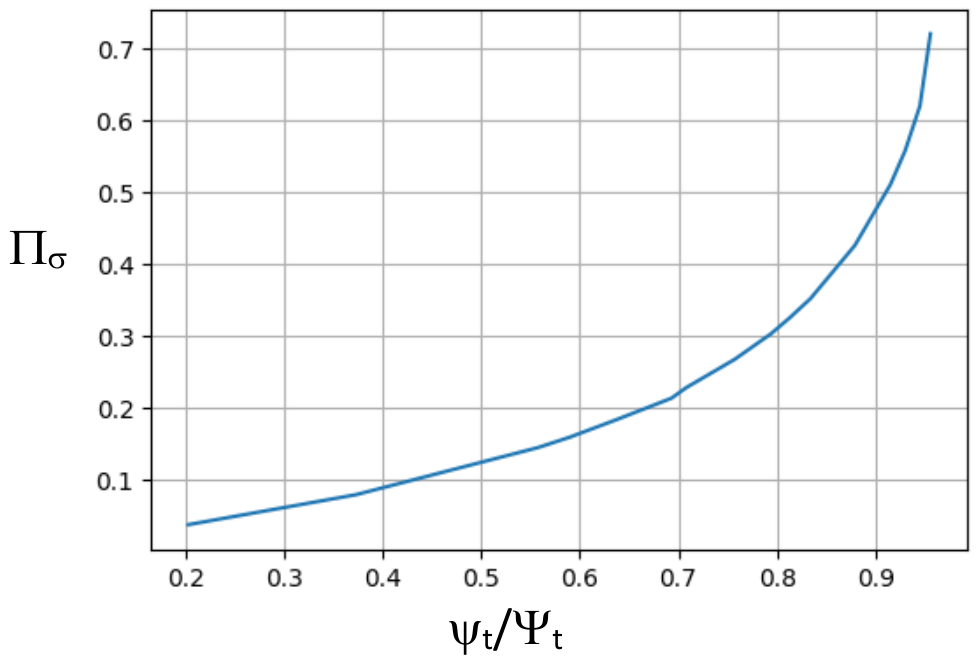}}
\caption{ The shape function has the form $\sigma=(a^2/\pi R^2)(\psi_t/\Psi_t) \Big(1+\Pi_\sigma(\psi_t/\Psi_t)\Big)$.  The function $\Pi_\sigma(\psi_t/\Psi_t)$ gives the effect of the separatrix, is defined in Equation (\ref{Pi-sigma}), and is given in the figure.} 
\label{fig:Pi_sigma}
\end{figure}

The toroidal magnetic flux enclosed by a magnetic surface is
\begin{eqnarray}
\psi_t(k) &=& \int_0^k \frac{\frac{d\psi_p^{in}}{dk} }{\iota(k)} dk \\
&=&\int_0^k \left(\frac{\mu_0 RI_p}{k}\right)\left(\frac{a^2 B_\varphi k^2 K(k)}{\mu_0RI_p}\right)dk\\
&=&a^2 B_\varphi \int_0^k k K(k) dk\\
&=&a^2 B_\varphi \Big(E(k) - (1-k^2)K(k)\Big)
\end{eqnarray} 
At $k=1$, the elliptic integral $E(1)=1$ and $K(k)$ goes to infinity logarithmically, so $\Psi_t= a^2 B_\varphi$ is the total toroidal magnetic flux enclosed by the separatrix.  

As $k\rightarrow0$, $E(k) - (1-k^2)K(k)\rightarrow \pi k^2/4$, and the ratio of $\psi_t$ to the total toroidal magnetic flux enclosed by the separatrix, $\psi_t/\Psi_t\rightarrow \pi k^2/4$ and $\sigma \rightarrow (a^2/\pi R^2) \psi_t/\Psi_t$.


\subsection{General expression for $\sigma(\psi_t/\Psi_t)$ \label{gen-sigma} }

The general expression for $\sigma(\psi_t)$, Equation (\ref{shape-function}) can be written as
\begin{eqnarray}
\sigma&=&\frac{a^2}{\pi R^2}\frac{\psi_t}{\Psi_t} \Big\{1+\Pi_\sigma(\psi_t/\Psi_t)\Big\}, \mbox{   where  }\\
\Pi_\sigma &=& \frac{1}{2}\frac{k^2 K}{E(k) - (1-k^2)K(k)}-1. \label{Pi-sigma}
\end{eqnarray}

Figure \ref{fig:Pi_sigma} illustrates the dependence of $\Pi_\sigma$ on $\psi_t/\Psi_t$.  ChatGPT recommended a fit to the data in Figure \ref{fig:Pi_sigma}:
\begin{equation} 
\Pi_\sigma\left(\frac{\psi_t}{\Psi_t}\right)=\frac{0.2019 \left(\frac{\psi_t}{\Psi_t}\right)^{1.1426}}{ \Big(1-\left(\frac{\psi_t}{\Psi_t}\right) \Big)^{0.4176}},
\end{equation}
which has a maximum error of less than 4\%.


\subsection{Location of 95\% of the poloidal flux in $\psi_t/\Psi_t$ \label{p95 in psi_t} }

\begin{figure}
\centerline{ \includegraphics[width=3 in]{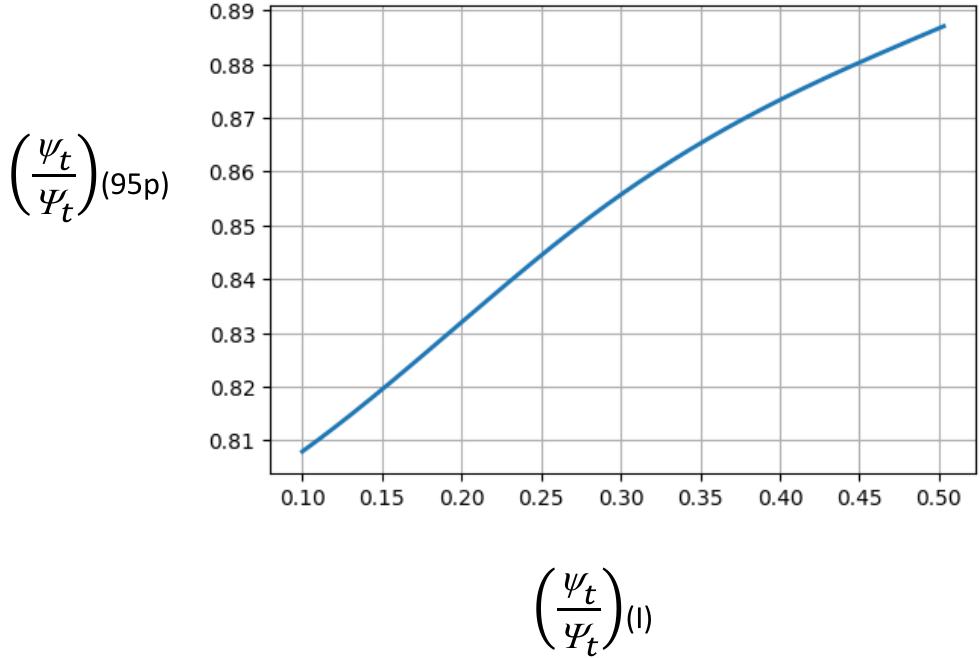}}
\caption{ The figure shows the value of $\psi_t/\Psi_t$ at which 95\% of the poloidal flux between the magnetic axis and the separatrix is reached as a function of the width of the current channel in the plasma $(\psi_t/\Psi_t)_{(I)}$.} 
\label{fig:psi_95p}
\end{figure}

The plasma current-density is assumed to have a flat profile out to $k_I$ and to be zero for larger $k$ with $k_I$ sufficiently small that the magnetic surfaces can be assumed to be circular.  With this assumption, the location in toroidal flux at which the current density drops to zero is $(\psi_t/\Psi_t)_{(I)} = \pi k_I^2/4$.  Equation (\ref{k_I k_95}), $k_{95p}=(k_I/\sqrt{e})^{0.05}$,  gives the relationship between $k_I$ and $k_{95p}$, the location in $k$-space at which 95\% poloidal flux between the axis and the separatrix is reached. 

Figure \ref{fig:psi_95p} gives the value of $(\psi_t/\Psi_t)=E(k)-(1-k^2)K(k)$ that corresponds to $k_{95p}$, which is denoted by $\psi_t^{p95}/
\Psi_t$.  Typically, 95\% of the poloidal flux corresponds to only 85\% of the toroidal flux.  


 \section{Summary \label{sec:summary} }

The toroidal magnetic flux $\psi_t$ enclosed by each magnetic surface is rarely used as a radial coordinate in the tokamak literature.  The preferred choice is the difference in the poloidal magnetic flux associated with a magnetic surface and that at the magnetic axis.  Faraday's Law says the slippage of the poloidal relative to toroidal flux is the loop voltage, Equation (\ref{flux-ev}).  It seems impossible to avoid confusions about the evolution of magnetic fields in a theory that ignores the toroidal flux.  

The focus on the poloidal flux in axisymmetric systems, such as tokamaks, comes from the Grad-Shafranov equation, which determines the poloidal flux as a function of position $\vec{x}$ without a direct reference to the toroidal flux.  Although the toroidal flux and $\Psi_p^{ax}(t)$ could be calculated, generally they are not.

A divertor defined by a separatrix is thought to be essential in tokamak power plants.  Research is required to determine what parameters can be used in tokamak power plants to ensure an acceptably low rate of disruptions.  As has been shown in this paper, both the edge safety factor and the internal inductance have a large but unnecessary definitional uncertainty when defined using the poloidal flux rather than the toroidal flux.   Defining the edge using a criterion based on 95\% of the poloidal flux corresponds to using approximately 85\% of the toroidal flux.  It is unclear what fraction of the toroidal flux is optimal for determining the disuptivity of tokamak plasmas.   

Artificial Intelligence (AI) could be used with large existing data sets from tokamak experiments as a basis for identifying parameters that ensure an acceptably low disruption probability.  Computational studies are of far lower cost than experiments for generating data and could be used to determine the linear tearing and external kink mode stability \cite{DCON:2016, DECON:2018} that could credibly arise during the plasma evolution of proposed tokamak power plants.  Remarkably, neither approach for identifying critical parameters is a focus of research.  For example, the recent paper on \emph{ARC disruption physics and strategy} \cite{ARC-dis} focused on disruption mitigation and survival.  A strategy for disruption avoidance was not discussed.

\section*{Acknowledgements}

This work received no external support.

 \vspace{0.01in}

\section*{Author Declarations}

The author has no conflicts to disclose. \vspace{0.01in}


\section*{Data availability statement}

Data sharing is not applicable to this article as no new data were created or analyzed in this study.


\appendix

\section{Jacobi elliptic functions and integrals \label{sec:elliptic}}

\begin{eqnarray}1&=&\cn^2(u) + \sn^2(u) \label{identity 0}\\
1&=& \dn^2(u) + k^2 \sn^2(u).   \label{identity 1} \\
\frac{d  \hspace{0.01in} \dn(u)}{du} &=& - k^2 \sn(u) \cn(u) \\ \nonumber\\
\frac{d \sn(u)}{du} &=&\cn(u)\dn(u) \\
\frac{d\cn(u)}{du} &=& - \sn(u)\dn(u)\\
\int_0^{4K} \dn(u) du &=& 2\pi \\
\int_0^{4K}\cn(u)du&=&0 \\
\int k K dk &=& E - K +k^2 K\\
&=& \frac{\pi k^2}{4} + \frac{\pi k^4}{32} +\frac{3\pi k^6}{256} + \cdots \\
K&=& \frac{\pi}{2} + \frac{\pi}{8}k^2 +\frac{9\pi}{128}k^4 + \cdots \\
E(1) &=& 1 \\
K(k\rightarrow1)& \rightarrow& \ln\Big(\frac{4}{\sqrt{1-k^2}}\Big).
\end{eqnarray}

Tables of the elliptic integrals $K(k)$ and $E(k)$ are given in Reference \cite{Elliptic Integrals} and on pages 608 and 609 of Reference \cite{Handbook}.



\begin{thebibliography}{99}


\bibitem{Boozer:Constraints2026} A. H. Boozer, \emph{Constraints on the magnetic field evolution in tokamak power plants}, Phys. Plasmas \textbf{33}, ??? (2026); doi:  10.1063/5.0322714; $\big<$https://arxiv.org/pdf/2507.05456$\big>$.

 \bibitem{Boozer:RMP} A. H. Boozer, \emph{Physics of magnetically confined plasmas}, Rev. Mod. Phys. \textbf{76}, 1071 (2004); doi: 10.1103/RevModPhys.76.1071.
 
 \bibitem{MHD stab} C. Z. Cheng, H. P. Furth and A. H. Boozer, \emph{MHD stable regime of the Tokamak}, Plasma Phys. Control. Fusion \textbf{29} 351 (1987); doi: 10.1088/0741-3335/29/3/006.

\bibitem{Eiditis:2021} N. Eidietis, \emph{Prospects for Disruption Handling in a Tokamak-based Fusion Reactor}, Fusion Sci. Technol. \textbf{77}, 732 (2021); doi 10.1080/15361055.2021.1889919.

\bibitem{Pedestal:2025} M.E. Fenstermacher, L.R. Baylor, E. de la Luna, M.G. Dunne, G.T.A. Huijsmans, A. Kirke, F.M. Laggner, T.H. Osborne, C. Paz-Soldan, S. Saarelma, P.B. Snyder, E. Viezzer, M. Becoulet, K.H. Burrell, A. Cathey, X. Chen, M. Hoelzl, J.W. Hughes, R. Maingi, A.O. Nelson, H. Urano, E. Wolfrum, X.Q. Xu, A. Diallo, L. Frassinetti, S. Futatani, L. Gil, R. Groebner, T. Happel, S.H. Kim, J. King, B. Labit, P.T. Lang, Y.Q. Liu, Z.X. Liu, R. Lunsford, G.Y. Park, U. Sheikh, W. Suttrop, B. Vanovac R.S. Wilcox, A. Wingen, and T. Zhang, \emph{Progress in pedestal and edge physics Chapter 3 of the special issue: on the path to tokamak burning plasma operation}, Nucl. Fusion \textbf{65},  053001 (2025); doi: 10.1088/1741-4326/adb1f3.

 \bibitem{JET-dis:2020} S.N. Gerasimov, P. Abreu, G. Artaserse, M. Baruzzo, P. Buratti, I.S. Carvalho, I.H. Coffey, E. De La Luna, T.C. Hender, R.B. Henriques, R. Felton, S. Jachmich, U. Kruezi, P.J. Lomas, P. McCullen, M. Maslov, E. Matveeva, S. Moradi, L. Piron, F.G. Rimini, W. Schippers, C. Stuart, G. Szepesi, M. Tsalas, D. Valcarcel1, L.E. Zakharov, and JET Contributors, \emph{Overview of disruptions with JET-ILW}, Nucl. Fusion \textbf{60}, 066028, (2020); doi: 10.1088/1741-4326/ab87b0.  
 
   \bibitem{Def-q_95} T. S. Taylor, E. J. Strait, L. Lao, A. G. Kellman, T. H. Osborne, K. Burrell, M. S. Chu, J. C. DeBoo, H. Fukumoto, P. Gohil, R. Groebner, C. Hsieh, G. Jackson, S. Kinoshita, P. Lomas, R. Snider, H. St. John, R. D. Stambaugh, R. E. Stockdale, and A. D. Turnbull, \emph{Achievement of Reactor-Relevant P in Lovv-q Divertor Discharges in the Doublet III-D Tokamak}, Phys. Rev. Lett. \textbf{62}, 1278 (1989); doi 10.1103/PhysRevLett.62.1278.
 
\bibitem{Boozer:NF-review} Allen H. Boozer, \emph{Non-axisymmetric magnetic fields and toroidal plasma confinement}, Nucl. Fusion \textbf{55},  025001 (2015); doi: 10.1088/0029-5515/55/2/025001.   

 \bibitem{Boozer:P-S} A. H. Boozer, \emph{Pfirsch-Schl\"uter Current} $<$https://arxiv.org/pdf/2605.21637$>$ (May 2026).
 
  \bibitem{Hegna-Callen:1994} C. C. Hegna and J. D. Callen, \emph{Stability of tearing modes in tokamak plasmas}, Phys. Plasmas \textbf{1}, 2308 (1994); doi: 10.1063/1.870628.
 
 \bibitem{La Haye:2006} R. J. La Haye, \emph{Neoclassical tearing modes and their control}, Phys. Plasmas \textbf{13}, 055501 (2006); doi: 10.1063/1.2180747.
 
 \bibitem{Fisch:1987} N. J. Fisch, \emph{Theory of current drive in plasmas}, Rev. Mod. Phys. \text{59}, 175 (1987); doi: 10.1103/RevModPhys.59.175.
 
\bibitem{Mag-coord} A. H. Boozer, \emph{ Plasma equilibrium with rational magnetic surfaces}, Phys. Fluids \textbf{24}, 1999 (1981); doi: 10.1063/1.863297.
 
 \bibitem{Landreman-Paul} M. Landreman and E. Paul, \emph{Magnetic Fields with Precise Quasisymmetry for Plasma Confinement}, Phys. Rev. Lett. \textbf{128}, 035001 (2022); doi: 10.1103/PhysRevLett.128.035001.
 
 \bibitem{Goodman} A. G. Goodman, K. C. Mata, S. A. Hennenberg, R. Jorge, M. Landreman, G. G. Plunk, H. M. Smith, R. J. J. Machenbach, C. D. Beidler. and P.  Helander, \emph{Constructing precisely quasi-isodynamic magnetic fields},  J. Plasma Phys. \textbf{ 89}, 905890504 (2023); doi: 10.1017/S002237782300065X.

 \bibitem{Hobbs-Taylor:1968} G. D. Hobbs, and J. B. Taylor, \emph{The Properties of Linear Filamentary Multipole Magnetic Fields}, UKAEA, Culham
Lab., Abingdon, Rep. CLM-R-95 (1968),  $<$https://scientific-publications.ukaea.uk/wp-content/uploads/CLM-R95.pdf$>$. 
 
  \bibitem{B-Rechester:1978} A. H. Boozer and A. B. Rechester, \emph{Effect of magnetic perturbations on divertor scrape-off width}, Phys. Fluids \textbf{21}, 682-689 (1978); doi: 10.1063/1.862277.
  
\bibitem{DCON:2016} A. H. Glasser,  Z. R. Wang, and  J.-K. Park, \emph{ Computation of resistive instabilities by matched asymptotic expansions}, Phys. Plasmas 23, 112506 (2016): doi 10.1063/1.4967862.
 
\bibitem{DECON:2018} A. S. Glasser and E. Kolemen, \emph{A robust solution for the resistive MHD toroidal $\Delta'$ matrix in near real-time} Phys. Plasmas \textbf{25}, 082502 (2018); doi: 10.1063/1.5029477.

\bibitem{ARC-dis} R. Sweeney, V. Riccardo, A. Braun, C. Clauser, A. J. Creely, T. Eich, I. Ekmark, A. Feyrer, C. Hansen, J. C. Hillesheim, T. Looby, S. Ratynskaia, R. Schramm, R. A. Tinguely, H. Wu, J. Bogusk, M. D. Boyer, J. Carmichael, A. Carter, R. Datta, T. F\"ul\"op, R. Granetz, S. Guizzo, M. Hoppe, A. LeViness, A. O. Nelson, K. Paschalidis, C. Paz-Soldan, I. Pusztai, C. Rea, T. Rizzi, A. R. Saperstein, P. B. Snyder, B. Stein-Lubrano. and P. Tolias, \emph{ARC disruption physics and strategy}, J. Plasma Phys. \textbf{92}, E68 (2026); doi: 10.1017/S0022377826101585.

\bibitem{Elliptic Integrals} J. M. Hammersley, \emph{Tables of Complete Elliptic Integrals}, Journal of Research of the National Bureau of Standards, \textbf{50}, 43 (1953).


\bibitem{Handbook} M. Abramowitz and I. A. Stegun \emph{Handbook of Mathematical Functions}, Dover Publications, New York, 1965) ISBN: 9780486612720.















 

 










\end{thebibliography}
 \end{document}